# Spectral Structures of Type II Solar Radio Bursts and Solar Energetic Particles

Running Head: Type II Solar Radio bursts and SEPs

Kazumasa Iwai[1*], Seiji Yashiro[2,3], Nariaki V. Nitta[4], and Yûki Kubo[5]

1. Institute for Space-Earth Environmental Research, Nagoya University, Furo-cho, Chikusa-ku, Nagoya, 464-8601, Japan
2. The Catholic University of America, Washington, DC 20064, USA
3. NASA Goddard Space Flight Center, Greenbelt, MD 20771, USA
4. Lockheed Martin Solar and Astrophysics Laboratory, Palo Alto, CA, USA
5. National Institute of Information and Communications Technology, 4-2-1 Nukui-kita, Koganei, Tokyo 184-8795, Japan

*Corresponding author email: k.iwai@isee.nagoya-u.ac.jp

Abstract

We investigated the relationship between the spectral structures of type II solar radio bursts in the hectometric and kilometric wavelength ranges and solar energetic particles (SEPs). To examine the statistical relationship between type II bursts and SEPs, we selected 26 coronal mass ejection (CME) events with similar characteristics (e.g., initial speed, angular width, and location) observed by the Large Angle and Spectrometric Coronagraph (LASCO), regardless of the characteristics of the corresponding type II bursts and the SEP flux. Then, we compared associated type II bursts observed by the Radio and Plasma Wave Experiment (WAVES) onboard the Wind spacecraft and the SEP flux observed by the Geostationary Operational Environmental Satellite (GOES) orbiting around the Earth. We found that the bandwidth of the hectometric type II bursts and the peak flux of the SEPs has a positive correlation (with a correlation coefficient of 0.64). This result supports the idea that the nonthermal electrons of type II bursts and the nonthermal ions of SEPs are generated by the same shock and suggests that more SEPs may be generated for a wider or stronger CME shock with a longer duration. Our result also suggests that considering the spectral structures of type II bursts can improve the forecasting accuracy for the peak flux of gradual SEPs.

# 1. INTRODUCTION

Solar eruptive phenomena, such as flares and coronal mass ejections (CMEs), generate high-energy particles called solar energetic particles (SEPs; e.g., Klein & Dalla 2017). Severe SEP events sometimes cause satellite anomalies and radiation exposure to humans in space (Cucinotta et al. 2010). Understanding and forecasting SEPs is an important issue in space weather.

Two types of SEP events include impulsive SEPs, which are associated with impulsive solar flares, and gradual SEPs, which are associated with CMEs (Reames 2013, and references therein). Gradual SEPs are primarily associated with energetic CMEs in the western hemisphere and are closely related to type II radio bursts (Kahler et al. 1984; Gopalswamy et al. 2008). These characteristics suggest that gradual SEPs are generated by shocks driven by CMEs in interplanetary space.

Type II solar radio bursts are nonthermal radio emissions with negative frequency drift observed between the metric and kilometric frequency range (McLean & Labrum 1985). They are thought to be plasma emissions generated by nonthermal electrons accelerated by shocks propagating in the corona and/or interplanetary space (e.g., Holman & Pesses 1983). Type II bursts in the decameter–hectometric (DH) and kilometric (km) range are called interplanetary (IP) type II bursts because the emission frequency corresponds to the typical plasma frequency of the outer corona and interplanetary space.

Although the IP type II bursts are generated from nonthermal electrons, while the SEPs are accelerated ions, many studies have suggested a close relation between the two phenomena (e.g., Cane & Stone 1984, Gopalswamy et al. 2005, 2008; Richardson et al. 2014; 2018; Winter & Ledbetter 2015). Additionally, some studies suggest that the high-energy electrons generating the radio bursts and SEPs are accelerated by the same shock (Gopalswamy

et al. 2018).

Previous studies on the relationship between SEPs and type II bursts investigated their associations. The spectral characteristics of type II bursts, such as the emission frequency, bandwidth, duration, and flux density, are thought to contain important information concerning the associated shock waves (e.g., Vršnak et al. 2001, 2002, 2004), but their relationship with the SEP characteristics are not well understood. The statistical relationship between the spectral characteristics of IP type II bursts and SEPs is important for understanding the generation mechanism of the SEPs via associated CME shocks. Fortunately, IP type II bursts are suitable for statistically investigating their spectral characteristics for two reasons. First, IP type II bursts can be continuously observed by a space-based radio instrument. Second, IP type II bursts at frequencies lower than 1 MHz are usually isolated from other types of solar radio bursts, such as type III and IV. Such advantages may not be applicable to metric type II bursts, even though they may contain important information on the acceleration of SEPs at shocks low in the corona (see section 4.3). Compared with IP type II bursts, we often find it hard to objectively isolate metric type II bursts from other types of emission, and so their spectral characteristics may not be derived on a statistical basis.

The purpose of this study is to understand the shock characteristics that generate the SEPs by considering IP type II bursts. We investigate the statistical relationship between the peak SEP flux density observed at 1 AU and the spectral characteristics of the corresponding IP type II bursts in the hectometric and kilometric ranges using selected CMEs with similar characteristics. Furthermore, we discuss the shock characteristics that generate SEPs from the type II bursts. The dataset and analysis methods used in this study are described in Section 2. The data analysis results are presented in Section 3 and discussed in Section 4. We summarize this study in Section 5.

## 2. METHODS

2.1. Data and Event Selection

More energetic CMEs can probably associate with type II bursts and SEP events (e.g., Gopalswamy et al 2008). Conversely, we investigate the relationship between the frequency characteristics of type II bursts and SEPs in this study. In order to avoid the bias of the big flare/CME syndrome (Kahler 1982a), we selected CMEs that have similar characteristics observed by the Large Angle and Spectrometric Coronagraph (LASCO; Brueckner et al. 1995) onboard the Solar and Heliospheric Observatory (SOHO) from the SOHO LASCO CME catalog[1] (Yashiro et al. 2004). The selection criteria are as follows: a velocity between 1200 km/s and 1800 km/s and an angular width > 120°. Because of the magnetic field connection governed by the Parker spiral, we expect higher likelihood to observe SEPs if the CME is from the western hemisphere. Therefore, we limit the range of longitude for the CME source region to be between 0° and 120°. The scale of the associated SEP events is defined as the peak of the >10-MeV proton intensity observed by the Geostationary Operational Environmental Satellite (GOES) at 1 AU. We exclude some CMEs that occurred during elevated SEP background period owing to the previous events. During the period of 2006 December to 2017 December, 26 CMEs met these criteria. Table 1 lists the 26 selected CME events. Even though those CME events have similar characteristics, the associated SEP flux at 1 AU has high variability, from 0 pfu to 1600 pfu.

---

[1] https://cdaw.gsfc.nasa.gov/CME_list/

Table 1. CME events and associated X-ray flares, SEPs, and type II bursts used in the statistical analysis in this study. BW, df/f, and flux refer to the bandwidth (kHz), bandwidth-to-frequency ratio, and flux density relative to the cosmic background ($10^{-6} V/\sqrt{Hz}$) of the burst components, respectively. "T.A." and "B.A." refer to "time-averaged" and "burst-averaged" spectral characteristics, respectively.

| CME | | | | SEP | X ray | | | H-Type-II | | | | | | km-Tpye-II | | | | | |
|---|---|---|---|---|---|---|---|---|---|---|---|---|---|---|---|---|---|---|---|
| Date | Time | Speed (km/s) | Width (deg) | GOES | Class | Location | B.A. BW (kHz) | T.A. BW (kHz) | B.A. df/f | T.A. df/f | B.A. flux | T.A. flux | B.A. BW (kHz) | T.A. BW (kHz) | B.A. df/f | T.A. df/f | B.A. flux | T.A. flux |
| 2006/12/13 | 2:54 | 1774 | 360 | 698 | X3.4 | S06W23 | 263 | 263 | 0.50 | 0.50 | 0.4 | 0.4 | 125 | 103 | 0.59 | 0.49 | 4.2 | 3.5 |
| 2010/8/14 | 10:12 | 1205 | 360 | 14 | C4.4 | N17W52 | 0 | 0 | 0.00 | 0.00 | 0.0 | 0.0 | 0 | 0 | 0.00 | 0.00 | 0.0 | 0.0 |
| 2010/8/18 | 5:48 | 1471 | 184 | 4 | C4.5 | N17W101 | 64 | 1 | 0.08 | 0.00 | 0.4 | 0.0 | 0 | 0 | 0.00 | 0.00 | 0.0 | 0.0 |
| 2011/6/7 | 6:49 | 1255 | 360 | 73 | M2.5 | S21W54 | 37 | 3 | 0.09 | 0.01 | 0.4 | 0.0 | 109 | 105 | 0.44 | 0.42 | 2.4 | 2.3 |
| 2011/8/4 | 4:12 | 1315 | 360 | 80 | M9.3 | N19W36 | 107 | 82 | 0.21 | 0.16 | 0.8 | 0.6 | 48 | 36 | 0.22 | 0.16 | 0.7 | 0.5 |
| 2011/8/8 | 18:12 | 1343 | 237 | 4 | M3.5 | N16W61 | 0 | 0 | 0.00 | 0.00 | 0.0 | 0.0 | 0 | 0 | 0.00 | 0.00 | 0.0 | 0.0 |
| 2011/8/9 | 8:12 | 1610 | 360 | 27 | X6.9 | N17W69 | 0 | 0 | 0.00 | 0.00 | 0.0 | 0.0 | 0 | 0 | 0.00 | 0.00 | 0.0 | 0.0 |
| 2012/3/18 | 0:24 | 1210 | 360 | 0 | ---- | N18W116 | 51 | 21 | 0.08 | 0.03 | 0.5 | 0.2 | 0 | 0 | 0.00 | 0.00 | 0.0 | 0.0 |
| 2012/5/17 | 1:48 | 1582 | 360 | 255 | M5.1 | N11W76 | 209 | 205 | 0.28 | 0.27 | 1.2 | 1.2 | 44 | 23 | 0.25 | 0.13 | 0.7 | 0.4 |
| 2012/6/23 | 7:24 | 1263 | 360 | 0 | C2.7 | N18W101 | 0 | 0 | 0.00 | 0.00 | 0.0 | 0.0 | 0 | 0 | 0.00 | 0.00 | 0.0 | 0.0 |
| 2012/7/8 | 16:54 | 1495 | 157 | 19 | M6.9 | S17W74 | 60 | 2 | 0.06 | 0.00 | 0.3 | 0.0 | 80 | 78 | 0.41 | 0.40 | 0.5 | 0.4 |
| 2012/7/19 | 5:24 | 1631 | 360 | 80 | M7.7 | S13W88 | 269 | 91 | 0.43 | 0.15 | 2.5 | 0.9 | 0 | 0 | 0.00 | 0.00 | 0.0 | 0.0 |
| 2013/5/22 | 13:25 | 1466 | 360 | 1660 | M5.0 | N15W70 | 446 | 446 | 0.97 | 0.97 | 2.8 | 2.8 | 157 | 147 | 0.60 | 0.56 | 0.7 | 0.7 |
| 2013/8/17 | 19:12 | 1202 | 360 | 0.7 | M1.4 | S05W30 | 96 | 31 | 0.15 | 0.05 | 0.6 | 0.2 | 58 | 14 | 0.20 | 0.05 | 0.5 | 0.1 |
| 2013/10/28 | 4:48 | 1201 | 156 | 4 | M5.1 | N08W71 | 429 | 387 | 0.69 | 0.62 | 6.6 | 6.0 | 8 | 1 | 0.04 | 0.00 | 0.3 | 0.0 |
| 2014/1/6 | 8:00 | 1402 | 360 | 42 | ---- | S15W112 | 214 | 115 | 0.30 | 0.16 | 1.2 | 0.6 | 0 | 0 | 0.00 | 0.00 | 0.0 | 0.0 |
| 2014/4/18 | 13:25 | 1203 | 360 | 59 | M7.3 | S20W34 | 207 | 68 | 0.24 | 0.08 | 2.1 | 0.7 | 0 | 0 | 0.00 | 0.00 | 0.0 | 0.0 |
| 2015/6/14 | 4:12 | 1228 | 195 | 0 | C5.9 | S12W34 | 0 | 0 | 0.00 | 0.00 | 0.0 | 0.0 | 0 | 0 | 0.00 | 0.00 | 0.0 | 0.0 |
| 2015/6/18 | 1:25 | 1714 | 195 | 17 | M1.2 | S16W81 | 0 | 0 | 0.00 | 0.00 | 0.0 | 0.0 | 0 | 0 | 0.00 | 0.00 | 0.0 | 0.0 |
| 2015/6/25 | 8:36 | 1627 | 360 | 16 | M7.9 | N09W42 | 210 | 210 | 0.70 | 0.70 | 0.6 | 0.6 | 0 | 0 | 0.00 | 0.00 | 0.0 | 0.0 |
| 2015/9/20 | 18:12 | 1239 | 360 | 3 | M2.1 | S20W24 | 78 | 6 | 0.12 | 0.01 | 0.4 | 0.0 | 77 | 76 | 0.37 | 0.36 | 0.7 | 0.7 |
| 2015/12/28 | 12:12 | 1212 | 360 | 4 | M1.8 | S23W11 | 0 | 0 | 0.00 | 0.00 | 0.0 | 0.0 | 89 | 57 | 0.28 | 0.18 | 1.0 | 0.7 |
| 2016/1/1 | 23:24 | 1730 | 360 | 22 | M2.3 | S25W82 | 150 | 65 | 0.22 | 0.10 | 1.5 | 0.6 | 0 | 0 | 0.00 | 0.00 | 0.0 | 0.0 |
| 2017/7/14 | 1:25 | 1200 | 360 | 22 | M2.4 | S06W29 | 80 | 4 | 0.10 | 0.01 | 0.5 | 0.0 | 161 | 161 | 0.76 | 0.76 | 4.4 | 4.4 |
| 2017/9/4 | 20:12 | 1418 | 360 | 210 | M5.5 | S10W12 | 0 | 0 | 0.00 | 0.00 | 0.0 | 0.0 | 0 | 0 | 0.00 | 0.00 | 0.0 | 0.0 |
| 2017/9/6 | 12:24 | 1571 | 360 | 844 | X9.3 | S08W33 | 224 | 224 | 0.44 | 0.44 | 1.5 | 1.5 | 122 | 122 | 0.66 | 0.66 | 0.9 | 0.9 |

2.2. Data Analysis

The IP type II burst data were observed by the Radio and Plasma Wave Experiment (WAVES; Bougeret et al. 1995) onboard the Wind spacecraft. In this study, we used 1-min-averaged Rad1 receiver data, which are observed in the frequency range between 20 kHz and 1040 kHz, whose data unit is in terms of ratio to the background. Figure 1 shows the radio dynamic spectra of the type II bursts on 2006 December 13 (up) and 2013 August 17 events (bottom). First, we identified type II radio emissions whose intensity is equal or more than 30% enhancement from the background level; they are indicated by the white contours in Figure 1. Weak radio emissions less than 30% enhancement were ignored. Subsequently, we removed the data periods that were contaminated by other radio emissions such as type III solar radio bursts and auroral kilometric radiations (AKR). For example, the type II burst observed on 2006 December 13 (Figure 1a) overlapped with AKRs after 4:00 UT, so we could not measure the lower frequency boundary of the type II. Such periods were excluded from the analysis. The upper and lower boundaries of the identified Type II burst components at a given time are indicated by the white rectangles in Figure 1.

Following Richardson et al (2014; 2018), this study investigated the IP type II bursts that were emitted below 1040 kHz (Rad 1 frequency band) because this radio emission frequency corresponds to the plasma frequency of the interplanetary space; therefore, they are more likely to be associated with the shocks of propagating CMEs. We investigated the type II characteristics in two frequency domains: one in the hectometric range (300–1040 kHz), which approximately corresponds to 7–25 solar radii ($R_s$) from the Sun, and the other in the kilometric range (lower than 300 kHz), which corresponds to 25 $R_s$ and above according to the density model of the interplanetary space (Erickson, 1964). Once the high-frequency side of the negatively drifting type II emission reached each observing frequency band, we began to derive the spectral characteristics of the type II component at any given time. The bandwidth and center frequency (indicated by the red crosses in Figure 1) of a given type II burst were derived from the difference between the highest and lowest frequencies of the burst component at a given time. Type II bursts sometimes have band-splitting structures (Vršnak et al. 2001). To derive the bandwidth of the type-II emission, we considered both the upper and lower bands of the band-splitting. Hence, the difference between the upper side of the upper band and lower side of the lower band should give the bandwidth (see Figure 1c). Conversely, harmonic emissions were carefully distinguished from the fundamental emission, and we only used the fundamental (or harmonic) emission when both of them are present. The bandwidth-to-frequency ratio was defined as the ratio between the bandwidth and the center frequency, as in Aguilar-Rodriguez et al. (2005). The peak flux was the largest flux density relative to the cosmic background between the highest and lowest frequencies of the type II component. Therefore, the peak flux was not the flux of the center frequency. The spectral characteristics of type II bursts (e.g. bandwidth and intensity) change in time, and it takes more than 60 minutes for the typical type II bursts of interest in this study to pass the hectometric band (300–1040 kHz). Therefore, we derived the averages of the type II characteristics from the first 60 minutes observations in each frequency domain (i.e., 60 one-minute-averaged spectra) except the contaminated time period. For some events, 60-min spectra could not be obtained because of contamination by other radio emissions. Herein, we used two types of averaging methods.

**Burst averaging**: IP type II bursts occasionally show spontaneous emissions and patch-like spectral structures; these bursts are shown in Figures. 1(c) and 1(d). We derived the averages from the period when the type-II burst emitted, implying that the time periods when the burst components were not observed were excluded from the averaging denominator. Notably, we used only the first 60-min spectra to detect the type II bursts for all the events. Therefore, averaging time is lower than 60 min in some patchy type II events. Hereafter, we call these

characteristics "the burst-averaged (B.A.) characteristics."

**Time averaging**: We averaged the entire time period regardless of the detection of the burst components. For example, the bandwidth at the time without type II emission is treated as 0 Hz. Therefore, the time-averaged spectral characteristics of the patchy type II bursts became smaller than that of burst-averaged. Hereafter, we call these "the time-averaged (T.A.) characteristics."

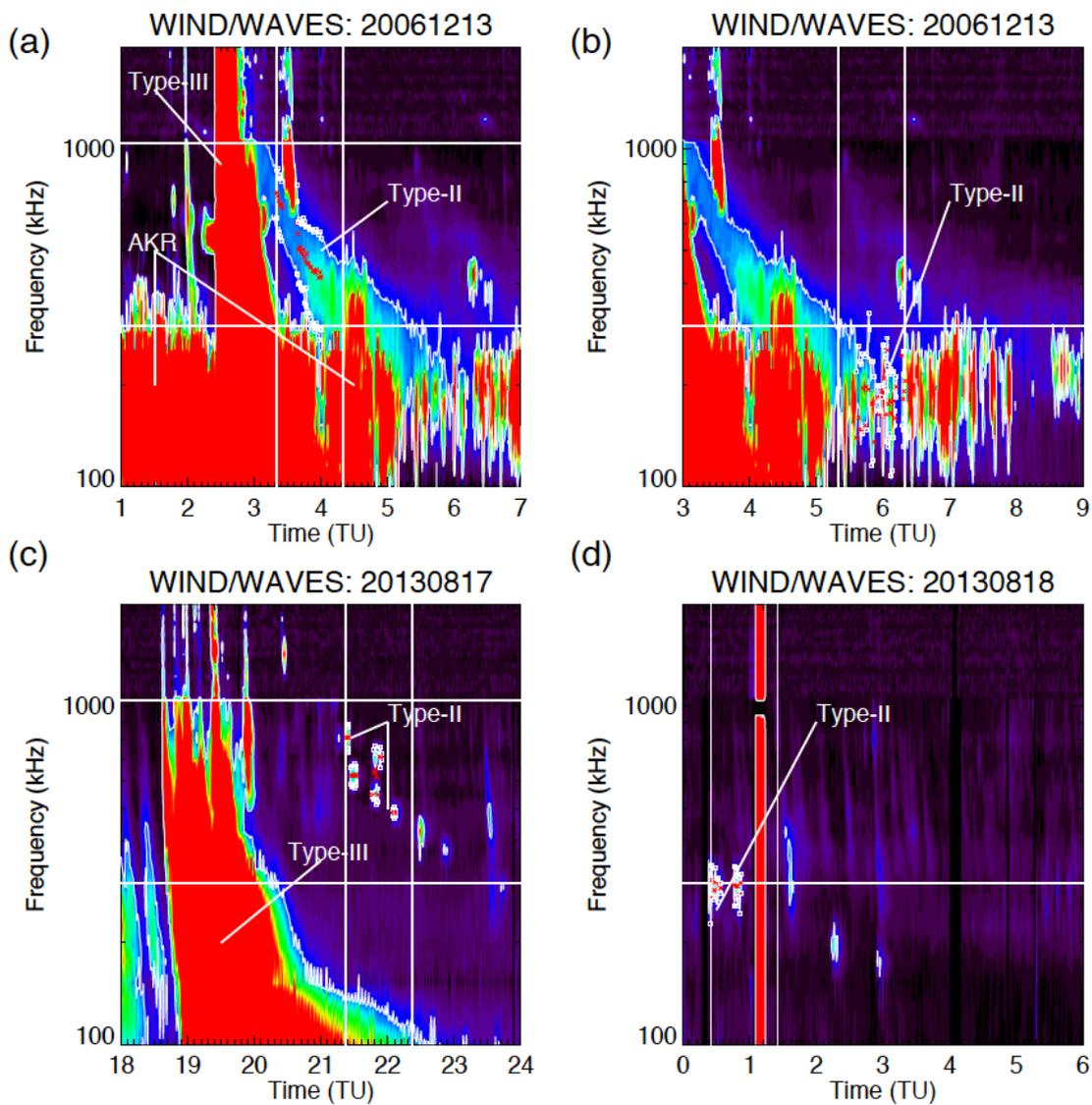

Figure 1. Radio dynamic spectra of type II bursts observed on 2006 December 13 with (a) hectometric and (b) kilometric type II bursts are highlighted. Radio dynamic spectra of type II bursts observed on 2013

August 17 and 18 with (c) hectometric and (d) kilometric type II bursts are highlighted. White contour: 30% enhancement from the background level. White rectangles: The highest and lowest frequencies of the type II component. Red crosses: The center frequency of the type II component. Vertical lines: The time period over which the spectral characteristics of the type II bursts are averaged. Horizontal lines: 300 and 1040 kHz that separate the data analysis bands in this study.

3. RESULTS

The top panel in Figure 2 shows the relationship between the CME speed and peak flux of SEP. For the events with no SEP enhancement, we plot them at 0.1 pfu for the display purpose. The regression line is determined by the least square method without no-SEP-enhancement events. Even though the CME speeds are limited in 1,200 – 1,800 km/s, there is a correlation to the SEP flux. Pearson's correlation coefficient (CC), $R_p$, is 0.53 and the Spearman's rank CC, $R_s$, is 0.47. The probability (P-value) of the observed (or more extreme) Spearman's rank CC by chance is 0.02. Therefore, the null hypothesis, that assumes the SEP flux has no relation to the CME speed, is ruled out with 95% confidence level. Cross and rectangle symbols indicate the events that B.A. bandwidth is equal or larger, and smaller than, 100 kHz, respectively. Out of the 10 large B.A. bandwidth events, 7 (or 70%) had larger SEP flux than the regression line. On the other hand, out of 15 small B.A. bandwidth events, only 4 (or 27%) had larger SEP flux. The middle and bottom panels in Figure 2 show the relationship between the CME speed and the B.A. bandwidth and the B.A. flux, respectively. In our limited number of samples, we could not find a significant relation between them.

The bottom panel in Figure 2 is a scatter plot between the CME speed and the B.A. radio flux in the hectometer domain. Pearson's and Spearman's CC are shown in the plot. No relation is found between them.

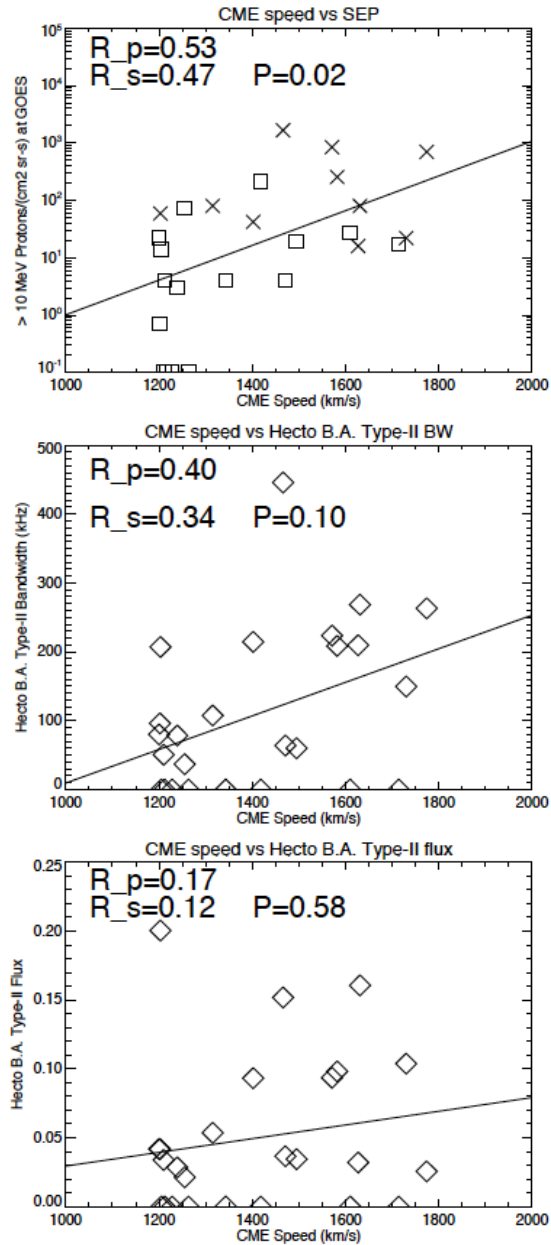

Figure 2. (Top) Scatter plot of the CME speed and SEP peak flux; rectangles and crosses show the events those B.A. bandwidths are smaller and larger than 100 kHz, respectively. (Middle) scatter plot of the CME speed and band-averaged (BA) hectometric type II bandwidth. (Bottom) scatter plot of the CME speed and BA hectometric type II flux. Pearson's CC ($R_p$), Spearman's rank CC, ($R_s$) and the P-value (P) are shown in each plot. The regression line is determined by the least squares method (solid line).

Figure 3 shows the relationship between the SEP fluxes and spectral characteristics of hectometric type II bursts. Note that we used only Rad 1 to detect hectometric type II bursts in this study. Therefore, DH type II burst emissions higher than 1040 kHz are missed, thereby reducing the number of CME events with IP type IIs. Eight CME events listed in Table 1 had no hectometric type-II bursts. They are included in the statistical analysis in Figures 2 and 3 as zero type II events (e.g., 0 kHz of bandwidth), although some events showed type II emissions above 1040 kHz.

There is one exceptional event observed on 2013 October 28, which is indicated by the "X" symbol in Figure 3. We excluded this event from the statistical analysis (i.e., the CCs) because the radio burst characteristics of this event was significantly affected by the CME–CME interaction, as discussed in Section 4.4.

The best correlation coefficient is derived from the B.A. and T.A. bandwidth ($R_p$= 0.64), which is shown in Figures 3a and 3b. The Spearman's rank CCs are 0.60 for both. The P-values are 0.0014 for the hectometric B.A. bandwidth and 0.0016 for the hectometric T.A. bandwidth. Therefore, the null hypothesis, which assumes the SEP flux has no relation to the type II bandwidths, is ruled out with 99% confidence level.

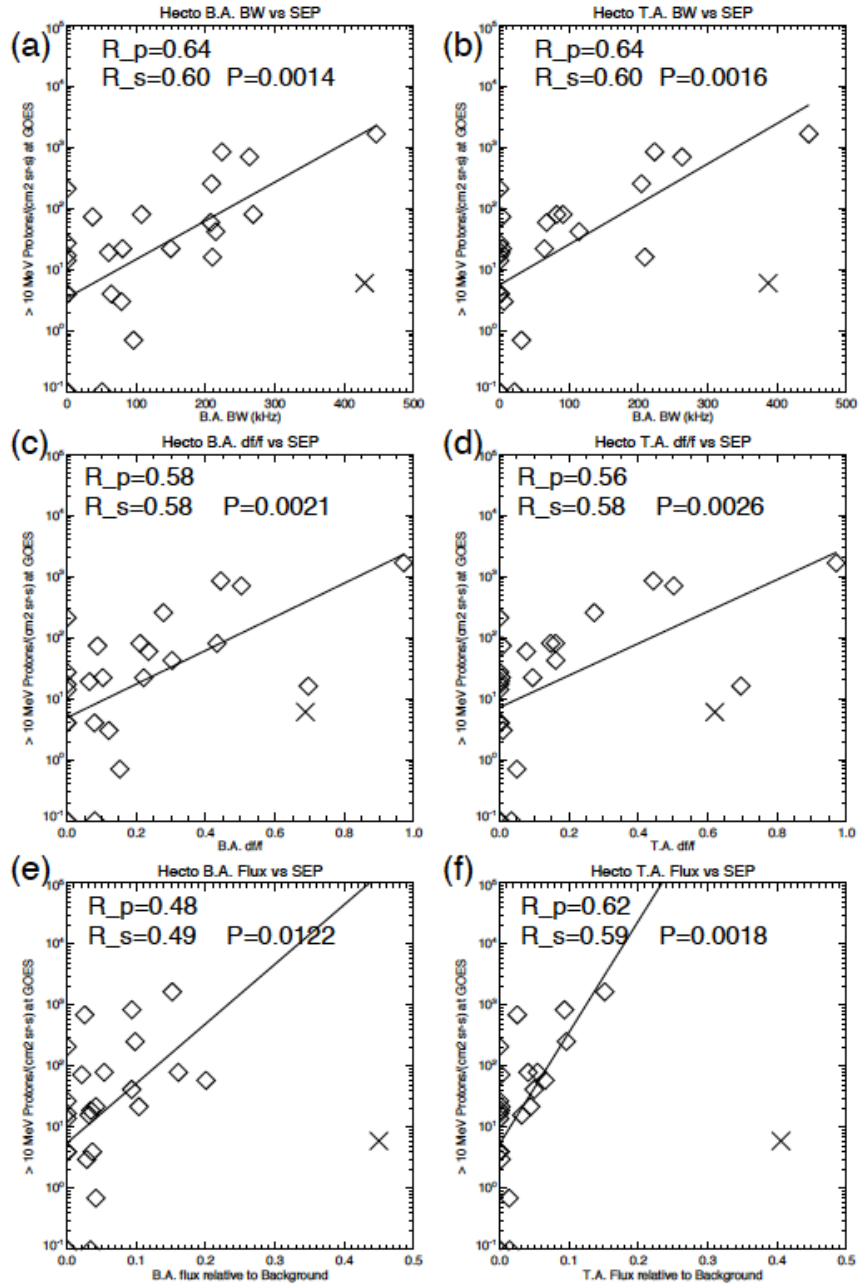

Figure 3. Scatter plots of the peak flux of SEPs and the spectral characteristics of hectometric type II bursts: (a) burst-averaged bandwidth; (b) time-averaged bandwidth; (c) burst-averaged bandwidth-to-frequency ratio; (d) time-averaged bandwidth-to-frequency ratio; (e) burst-averaged flux; and (f) time-averaged flux. The linear least squares fit lines and the CCs are shown for each plot. X indicates the CME–CME interaction event that was excluded from the statistical analysis.

Figure 4 shows the same scatter plots as Figure 3 but for type II bursts in the kilometric range. Even though there are weak correlations between SEPs and the spectral characteristics, their CCs are weaker than those in the hectometric band.

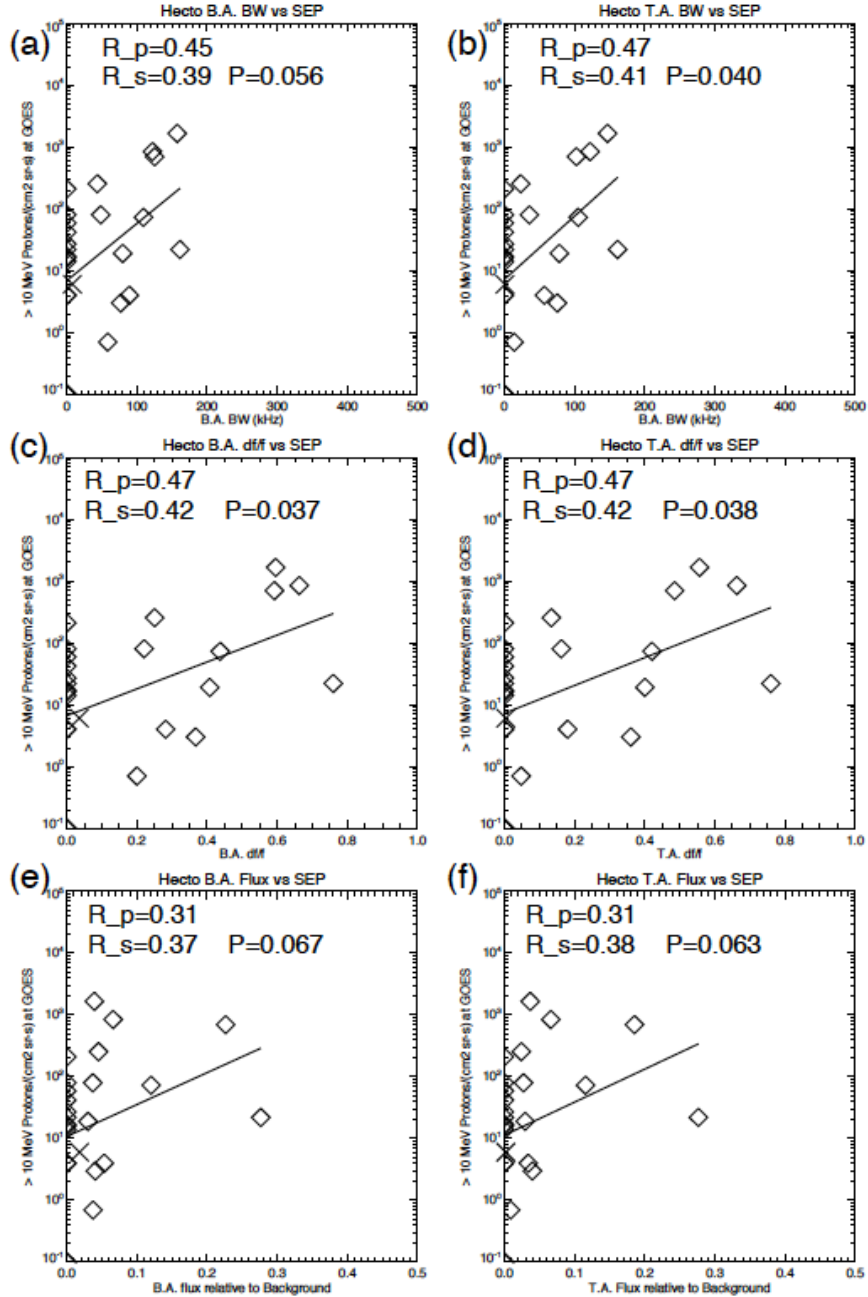

Figure 4. Same scatter plots as in Figure 3 for the spectral characteristics of kilometric type II bursts and the peak flux of SEPs.

## 4. DISCUSSION

In this study, we investigated type II bursts at frequencies lower than 1040 kHz, although other types of solar radio bursts are also associated with SEPs. For example, the so-called type III-l bursts, i.e. low frequency type III bursts that appear in the late phase of flares and last long, are often good indicators of SEP events (Cane et al. 2002; MacDowall et al. 2003). Additionally, Kahler (1982) showed SEP events associated with type IV bursts. However, type II burst more directly signify CME-driven shock waves that are responsible for gradual SEP events. Type II bursts at higher frequencies may also be correlated with SEP events, but Gopalswamy et al. (2008) showed that metric type II bursts are more likely associated with SEPs if they are also seen as IP type II bursts. Accordingly, the following discussions limit to the frequency characteristic of IP type II bursts.

### 4.1. Bandwidth and SEPs

In Figure 3, both T.A. and B.A. bandwidth of the hectometric type II bursts correlates well with the peak flux of SEP. The two correlation coefficients are similar values ($R_p = 0.64$, and $R_s = 0.60$) and the P-value of $R_s$ is very small ($P = 0.001 \sim 0.002$) for both T.A. and B.A. bandwidth, suggesting high significance of the derived correlation. From these results, we can conclude that the correlations between the bandwidth of the type II bursts and peak flux of SEP should be significant regardless of the analysis method of the spectral data and statistical analysis methods. In the top panel of Figure 2, most CMEs with the B.A. bandwidth larger than 100 kHz (x-marks) are indicated above the regression line, while more than a half of the CMEs with the B.A. bandwidth smaller than 100 kHz (rectangles) are indicated below the regression line. This result also suggests that the CMEs associated with type II bursts characterized by larger B.A. bandwidth tend to generate more SEPs. From these results, we consider the bandwidth of the type II bursts at this frequency band to hold important information on the generations of SEPs.

The bandwidth of the plasma emissions may correspond to the density difference in the emission region, and the density difference may represent the spatial scale of the emission region under the gradual density distribution of the inner heliosphere. Therefore, a wider bandwidth could correspond to a larger emission region. The positive correlation between the type II bandwidth and the SEPs may suggest that, if the particle acceleration occurs on a larger spatial scale, more SEPs could be generated. Additionally, a wider bandwidth is explained via the larger density gradient at the radio source region. Such region can be more turbulent and may lead more efficient particle acceleration.

Another interpretation is possible using a model proposed by Vršnak et al. (2001) in which the band-splitting width of type II bursts corresponds to the density gap in the shock. In this model, the Alfvén Mach number of the shock can be estimated from the bandwidth of the splitting by assuming the density gradient along the shock trajectory. The correlation between the type II bandwidth and the SEP flux derived in this study may suggest that a shock with a larger Alfvén Mach number can produce more SEP particles, even though there are type II bursts without clear band-splitting on our list. Note that another theory has been proposed, which appears to contradict this scenario (Du et al. 2014, 2015).

4.2. Radio Flux and SEPs

In Figures 3(e) and 3(f), the B.A. radio flux of hectometric type II bursts and SEPs have a relatively low correlation ($R_p$ = 0.48); however, the correlation becomes better after time averaging ($R_p$ = 0.62). Solar radio bursts generated by plasma emission contain many physical processes, such as plasma wave generation, radio wave emission, and propagation (e.g., Li et al. 2008). The modulation of the spectral structures of type II bursts can be explained by the modulation of the radio emission and the propagation processes along the trajectory of the shock (Schmidt and Cairns 2012, 2016). Therefore, a strong radio emission does not necessarily mean a strong shock, and it is not surprising that the peak flux of a type II burst has a low correlation with SEPs. Conversely, the better correlation with SEPs after time averaging suggests that the radio burst emission time may correspond to the acceleration time of the SEPs.

In Figure 2, the relationships between the CME speed and the type II spectral parameters (the burst-averaged bandwidth and the burst-averaged flux) are weak. Nevertheless, the spectral parameters of type II bursts are well correlated with SEPs. Therefore, we suggest that the electron beams of type II bursts and SEPs are generated by the same particle acceleration processes, but are controlled by something other than the CME speed.

The emission processes of IP type II bursts are beyond the scope of this study. Some studies have even suggested that wide-band IP type II emissions might be synchrotron emissions from trapped electrons within the CME (Bastian 2007, Pohjolainen et al. 2013). Different radio emission processes can excite different radio fluxes from the same amount of high-energy particles.

4.3. Radio emission frequency range and SEPs
Spectral characteristics of type II bursts in the kilometric range have lower correlations with SEPs compared to those in the hectometric range, as shown in Figures 3 and 4. Fast CMEs, such as the ones accommodated in this study, tend to decelerate in the LASCO field of view, suggesting that the shock waves driven by them become weaker at greater distances from the Sun. That may suggest that these shock waves have higher particle acceleration efficiencies in regions closer to the Sun. For well-connected high-energy SEP events at hundreds of MeV, the first arriving particles often correspond to CME heights of a few $R_s$ (Reames 2009), and the SEP peaks are reached when

CMEs are at heights of 5-15 $R_s$ (Kahler 1994). SEPs at lower energies such as >10 MeV may have similar corresponding CME heights unless the time profiles are dominated by energetic storm particles (ESPs). Therefore, we may expect better correlations of spectral characteristics of type II bursts at wavelengths shorter than the kilometric range (see Figures 3 and 4). There is another possibility that explains the lower collation between the frequency characteristic of the kilometric type II bursts and SEPs. The plasma frequency of the hectometric range in this study (300–1020 kHz) corresponds to up to 0.1 AU from the solar surface. Therefore, the distance between the radio source region and the Earth should be approximately 0.9–1.0 AU regardless of the location of the radio source region. Conversely, plasma emission in the kilometric range corresponds to 0.1–1 AU. The distance between the Earth and the emission region can vary significantly depending on the longitudinal difference between the Earth and the radio source region. This likely explains the lower correlation between the kilometric flux and the SEPs.

Given that the CME-driven shock is stronger closer to the Sun, it is possible that type II bursts at higher frequencies than sampled by the RAD1 receiver of Wind/WAVES, that is above 1 MHz, may be equally relevant for SEP studies. We therefore examined the same set of type II bursts observed by the RAD2 receiver, which covers 1-14 MHz. However, we found it difficult to conduct the same analysis described in section 2.2, largely because we would need the same type II bursts to be observed in the metric range so that we could define the upper frequencies for the bandwidth. In many cases, this was apparently impossible, since the radio dynamic spectra tend to be complex in the metric range with the presence of type III (and IV) bursts observed in overlapping times and frequencies with type II bursts (e.g. Nitta et al. 2014). We also note that a few type II bursts in our sample are hardly noticeable in the 1-14 MHz range. These observational constraints have unfortunately prevented us from statistically studying the spectral characteristics of type II bursts above 1 MHz. On the other hand, reasonably high correlations of the bandwidth of hectometric type II bursts with the SEP peak flux has established the potential importance of this frequency range in predicting the magnitude of SEP events.

4.4. CME–CME Interaction on 2013 October 28

Figure 5 shows a white-light coronagraph image and the radio dynamic spectra on 2013 October 28. Two CMEs (CME1 and CME2) occurred, which are indicated by the black arrows. The white vertical line in the right panel indicates the time when the coronagraph image in the left panel was taken. It appears that the type II burst appeared to be significantly enhanced immediately after the CME–CME interaction began. Type II enhancements during CME–CME interactions have been reported in multiple studies (e.g., Gopalswamy et al. 2002, Al-Hamadani et al. 2017). Pohjolainen et al. (2016) reported that some CME–CME interactions do not generate SEP enhancements. They explained that the earlier CMEs and shocks change the propagation paths or prevent the propagation of SEPs because the type III bursts observed after a CME–CME interaction stop at much higher frequencies than those at the earlier events. As shown in the left panel in Figure 5, type III radio bursts around 7-9 UT stopped at much higher frequencies than the earlier type III bursts. This behavior can be explained by the hypothesis proposed by Pohjolainen et al. (2016). In addition, the type III bursts also stopped at higher frequencies than the simultaneous type II bursts, suggesting a lack of open field lines beyond the shock. Alternatively, the stop of type III bursts can be related to the change of excitation condition of plasma emission and/or radio emission due to the CME–CME interaction. That situation may also prevent to derive the relationship between the type II radio bursts and SEPs. Therefore, we removed this event from the statistical analysis although it is still listed in Table 1. Note that the kilometric counterpart of this type II burst event was not significant probably because the CME–CME interaction had already finished when the CME reached the plasma emission region of kilometric bursts (>70$R_s$).

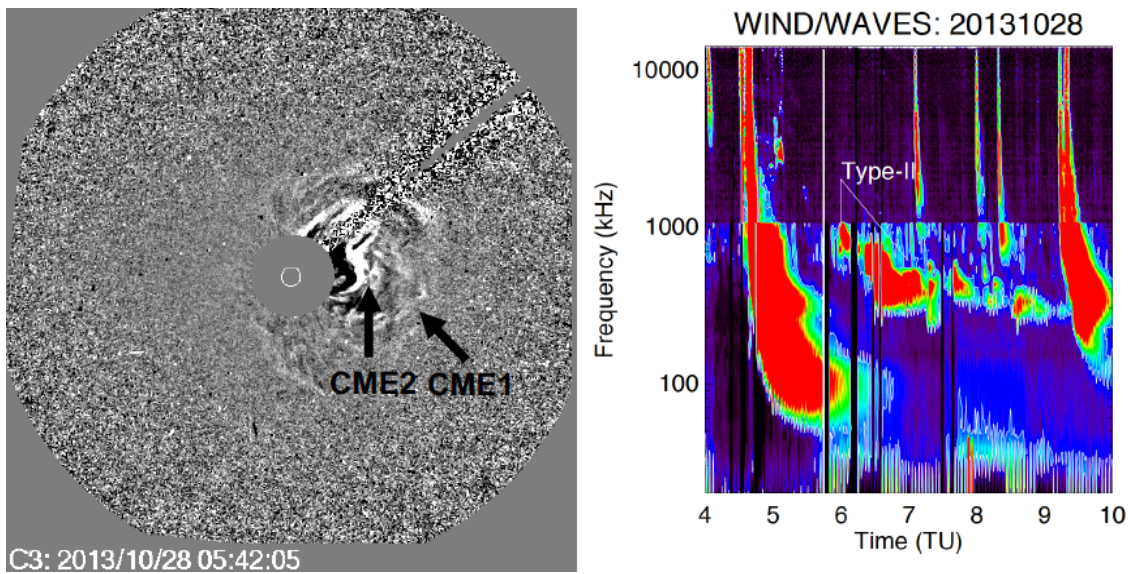

Figure 5. (left) Difference image of a white-light coronagraph obtained by SOHO/LASCO on 2013 October 28. The two CMEs are indicated by the black arrows. (Right) Radio dynamic spectra from Wind/WAVES. The white vertical line indicates the time when the image in the left panel was obtained.

4.5. SEP events on 2017 September 4

Another exceptional event in our list is the SEP event on 2017 September 4. In Table 1, this event might be recognized as the SEP event without any type II bursts. Figure 6 shows the radio dynamic spectra for 2017 September 4. Numerous hectometric type III bursts in the frequency range below 1040 kHz were observed during the time when a type II burst should have been observed. Although some emissions similar to type II were observed in that band, contamination by type III bursts prevented the detection of type II burst emission. Note that type II emissions were observed at frequencies above 1040 kHz.

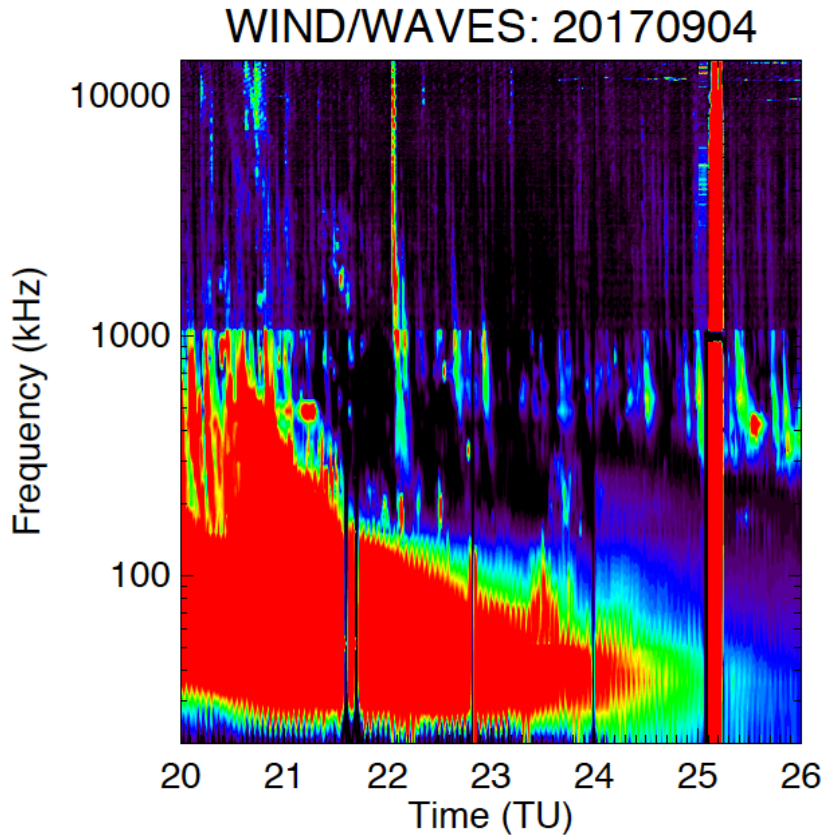

Figure 6 Radio dynamic spectra from Wind/WAVES observed on 2017 September 4.

## 5. CONCLUSIONS

We investigated the relationships between the spectral characteristics of IP type II bursts and SEPs. We selected 26 CME events that have similar CME characteristics (i.e., initial speed, angular width, and location) regardless of the corresponding type II bursts and SEP flux. This event selection enabled us to examine the statistical correlation between type II bursts and SEPs. The results of the statistical analysis are summarized as follows.

- The bandwidth of hectometric type II bursts and the peak flux of SEPs have a positive correlation ($R_p = 0.64$). This result supports the idea that the electron beam of a type II burst and the nonthermal ions of SEPs are generated by the same shock and suggest that more SEPs can be generated by a wider or stronger shock with a longer duration.

- The flux of a hectometric type II burst is also correlated with the SEPs even though this CC is lower than that of the bandwidth. This lower correlation may be caused by radio emission and/or propagation processes.
- The same spectral characteristics of type II bursts in the kilometric range have a lower correlation with SEPs than those in the hectometric range. This lower correlation may be caused by the weakening the CME shocks, or distance variation between the source region and the Earth.

Our result suggests that the spectral structures of type II bursts can be used to improve the forecasting accuracy for SEPs. In particular, the peak flux of gradual SEPs can be estimated more accurately than before with the spectral characteristics of type II bursts. Further studies of the same set of events might extend our results. For example, the shock scale and duration can be investigated via multiple white-light coronagraph observations using both the Wind and Solar TErrestrial RElations Observatory (STEREO) satellites. Propagating shocks outside the field-of-views of the coronagraphs can be traced using ground-based radio observations of interplanetary scintillation (e.g. Iwai et la 2019). Other frequency characteristics, such as start and end frequency of type II bursts, can provide additional information of CMEs and SEPs (e.g. Vasanth et al 2015).

Acknowledgments
This study is based on the results obtained from the Coordinated Data Analysis Workshop held in 2018 August 6-9 organized by the Project for Solar-Terrestrial Environment Prediction (PSTEP). This study was supported by MEXT/JSPS KAKENHI Grant Number 18H04442. The work of NVN was supported by NASA grant 80NSSC18K1126. This work was carried out by the joint research program of the Institute for Space-Earth Environmental Research (ISEE), Nagoya University. This work benefited from NASA's open data policy in using Wind and SOHO data, and NOAA's

GOES X-ray and particle data.